\documentclass[twocolumn]{autart}
\usepackage{graphicx,amstext,amsmath,amssymb,psfrag,enumerate,color}
\usepackage{parskip}
\usepackage{soul}
\newtheorem{ass}{Assumption}

\newtheorem{props}{Proposition}
\newtheorem{lemma}{Lemma}
\newtheorem{rmk}{Remark}

\newcommand{\arctanh}{\operatorname{arctanh}}

\newcommand{\dom}{\ensuremath{\text{dom}\,}} 

\begin{document}
\begin{frontmatter}

\title{Event-triggered control of nonlinear singularly perturbed systems based only on the slow dynamics\thanksref{footnoteinfo}}
\thanks[footnoteinfo]{The material of this paper was partially presented at the 9th IFAC Symposium on Nonlinear Control Systems (NOLCOS 2013), Toulouse, France. This work was partially supported by the ANR under the grant COMPACS (ANR-13-BS03-0004-02).}
\author{Mahmoud Abdelrahim}\ead{othmanab1@univ-lorraine.fr},    
\author{Romain Postoyan}\ead{romain.postoyan@univ-lorraine.fr},               
\author{Jamal Daafouz}\ead{jamal.daafouz@univ-lorraine.fr}  

\address{Universit\'{e} de Lorraine, CRAN, UMR 7039 and CNRS, CRAN, UMR 7039, France}  

\begin{keyword}                                                                                                 
Event-triggered, Singular Perturbation, Hybrid Systems.               
\end{keyword}                                                                                                   

\begin{abstract}
  Controllers are often designed based on a reduced or simplified model of the plant dynamics. In this context, we investigate whether it is possible to synthesize a stabilizing event-triggered feedback law for networked control systems (NCS) which have two time-scales, based only on an approximate model of the slow dynamics. We follow an emulation-like approach as we assume that we know how to solve the problem in the absence of sampling and then we study how to design the event-triggering rule under  communication constraints. The NCS is modeled as a hybrid singularly perturbed system which exhibits the feature to generate jumps for both the fast variable and the error variable induced by the sampling. The first conclusion is that a triggering law which guarantees the stability and the existence of a uniform minimum amount of time between two transmissions for the slow model may not ensure the existence of such a time for the overall system, which makes the controller not implementable in practice. The objective of this contribution is twofold. We first show that existing event-triggering conditions can be adapted to singularly perturbed systems and semiglobal practical stability can be ensured in this case. Second, we propose another technique that combines event-triggered and time-triggered results in the sense that transmissions are only allowed after a predefined amount of time has elapsed since the last transmission. This technique has the advantage, under an additional assumption, to ensure a global asymptotic stability property and to allow the user to directly tune the minimum inter-transmission interval. We believe that this technique is of its own interest independently of the two-time scale nature of the addressed problem. The results are shown to be applicable to a class of globally Lipschitz systems.
\end{abstract}
\end{frontmatter}
\section{Introduction} \label{itroduction}

The increasing popularity of embedded systems and networked control systems has motivated the development of new implementation paradigms in order to handle the resources limitations of these systems. Indeed, although periodic sampling is appealing from the analysis and implementation point of view, it may yield a conservative solution as it may unnecessarily use the network. Event-triggered control has been proposed as an alternative where it is the occurrence of an event, typically a variation of the plant state and not a clock, which closes the feedback loop \cite{Arzen1999simple}, \cite{Astrom1999comparison}. This may allow to significantly reduce the utilization of the resources compared to the periodic implementation, see e.g. \cite{Donkers2012output}, \cite{Heemels2012Introduction}, \cite{Romain2011unifying}, \cite{Tabuada2007event}, \cite{Wang2011Event}. Available techniques rely on the knowledge of an accurate model of the plant (which may be affected by uncertainties or external disturbances). However, the controller is often designed based on a \emph{reduced} or \emph{simplified} model of the plant dynamics. For two time-scale systems for instance, singular perturbation theory can be used to approximate the slow and the fast dynamics, see \cite{Khalil}, \cite{Kokotovic}. In this context, it is possible to design the controller based only on the slow model, when the origin of the fast model is asymptotically stable, for stabilizable linear time-invariant (LTI) systems \cite{Kokotovic}, classes of nonlinear systems (see Section 5.4 in \cite{Khalil}) and linear time-varying sampled data systems with periodic sampling \cite{Pan1994Hinf}. In this paper, we investigate whether this approach is applicable for event-triggered control.

We consider the scenario where the controller communicates with a two-time scale nonlinear system via a digital communication channel. Our objective is to design a stabilizing event-triggered feedback law based only on an approximate model of the slow dynamics. This problem is motivated by the fact that engineers often neglect the fast stable dynamics in practice and design the feedback law based only on the slow model. To the best of our knowledge, this is the first paper in that direction.

We cast the overall problem as a hybrid singularly perturbed system with the formalism of \cite{Teel}. The stability of this type of systems is analysed in \cite{Sanfelice2011On}, \cite{Wang2012analysis}, \cite{Wang2012averaging}. In this study, we address a design problem as we construct the flow and jump sets (i.e. the triggering condition) and we propose different stability analyses under a different set of assumptions. We highlight a specific challenge which arises with the event-triggered implementation: the state of the fast model experiences a jump at each transmission due to the change of variables we introduce to separate the slow and the fast dynamics using singular perturbation theory. These jumps induce non-trivial difficulties in the stability analysis. That is a feature of the problem which is not present in available results on event-triggered control where only the sampling-induced error is reset to zero at each transmission, see e.g. \cite{Donkers2012output}, \cite{Heemels2012Introduction}, \cite{Romain2011unifying}, \cite{Tabuada2007event}, \cite{Wang2011Event}.

We follow an emulation-like approach to design the event-triggered controllers (see \cite{Romain2011unifying}, \cite{Tabuada2007event}). We first synthesize a stabilizing controller for the approximate slow model obtained by singular perturbation theory, in the absence of communication constraints. Afterwards, we take into account the effect of the network and we design the event-triggering condition. The first observation we make is that, even if the triggering law guarantees the asymptotic stability of the origin of the slow model and the existence of a strictly positive lower bound on the inter-transmission times, such a time is no longer guaranteed to exist for the overall system. As a consequence, the controller is not implementable in practice. We then propose two classes of event-triggered controllers which overcome this issue. The first policy relies on the event-triggering conditions \cite{Donkers2012output}, \cite{Mazo2012decentralized} but it requires to fully modify the stability analysis to handle the features of the problem due to the two-time scale nature of the system. We show that a semiglobal practical stability property holds where the adjustable parameter appears in the event-triggering condition. The second technique combines the event-triggered implementation of \cite{Tabuada2007event} with the time-triggered results in \cite{Nesic2009explicit}, like in \cite{Mazo2011decentralized}, \cite{Tallapragada2013decentralized}, \cite{Tallapragada2012event}, \cite{Wang2012asynchronous}, in the sense that transmissions are only allowed after a predefined amount of time has elapsed since the last transmission. This allows us to directly tune the minimum transmission interval. We show that a global asymptotic stability property is satisfied in this case, under an additional assumption. The results are shown to be applicable to a class of globally Lipschitz systems, which include stabilizable LTI systems as a particular case.

The remainder of the paper is organised as follows. The problem is stated in Section \ref{sec: prob-statment}. The main assumptions are presented in Section \ref{subsec: assumptions}. In Section \ref{sec: main-results}, we state the main results. In Section \ref{sec: application-glob-Lips}, we show that the proposed event-triggered control strategies are applicable to a class of globally Lipschitz systems. The proofs are given in the Appendix.

\section{Preliminaries}
We denote $\mathbb{R} = (-\infty,\infty)$, $\mathbb{R}_{\geq 0} = [0,\infty)$, $\mathbb{Z}_{\geq 0} = \{ 0, 1, 2, . . \}$. The Euclidean norm is denoted as $|.|$. We use the notation $(x,y)$ to represent the vector $[x^{T}, y^{T}]^{T}$ for $x \in \mathbb{R}^{n}$ and $y \in \mathbb{R}^{m}$. A continuous function $\gamma: [0,\infty) \rightarrow \mathbb{R}_{\geq 0}$ is of class $\mathcal{K}$ if it is zero at zero, strictly increasing, and it is of class $\mathcal{K}_{\infty}$ if in addition $\gamma(s) \rightarrow \infty$ as $s \rightarrow \infty$. A continuous function $\gamma: \mathbb{R}_{\geq 0} \times \mathbb{R}_{\geq 0} \rightarrow \mathbb{R}_{\geq 0}$ is of class $\mathcal{KL}$ if for each $t \in \mathbb{R}_{\geq 0}$, $\gamma(.,t)$ is of class $\mathcal{K}$, and, for each $s \in \mathbb{R}_{\geq 0}$, $\gamma(s,.)$ is decreasing to zero. We denote the minimum and maximum eigenvalues of the symmetric positive definite matrix $A$ as $\lambda_{\min}(A)$ and $\lambda_{\max}(A)$ respectively. We will consider locally Lipschitz Lyapunov functions (that are not necessarily differentiable everywhere), therefore we will use the generalized directional derivative of Clarke which is defined as follows. For a locally Lipschitz function $V: \mathbb{R}^{n} \rightarrow \mathbb{R}_{\geq 0}$ and a vector $\upsilon \in \mathbb{R}^{n}$, $V^{\circ}(x;\upsilon) := \lim \sup_{h \to 0^{+}, \, y \to x}(V(y+h\upsilon) - V(y))/h$. For a $C^{1}$ function $V$, $V^{\circ}(x;\upsilon)$ reduces to the standard directional derivative $\langle\nabla V(x), \upsilon\rangle$, where $\nabla V(x)$ is the (classical) gradient. We will use the following result which corresponds to Proposition 1.1 in \cite{Nesic2008Lyapunov}.
\begin{lemma} \label{lma: clarke}
Consider two continuously differentiable functions $U_{1}: \mathbb{R}^{n} \rightarrow \mathbb{R}$ and $U_{2}: \mathbb{R}^{n} \rightarrow \mathbb{R}$. Let $A:=\{ x: U_{1}(x) > U_{2}(x) \}$, $B:=\{ x: U_{1}(x) < U_{2}(x) \}$ and $\Gamma:= \{ x: U_{1}(x) = U_{2}(x) \}$. For any $\upsilon \in \mathbb{R}^{n}$, the function $U: x \mapsto \max\{U_{1}(x), U_{2}(x)\}$ satisfies $U^{\circ}(x; \upsilon) = \langle \nabla U_{1}(x), \upsilon\rangle$ for all $x\in A$, $U^{\circ}(x; \upsilon) = \langle \nabla U_{2}(x), \upsilon\rangle$ for all $x\in B$ and $U^{\circ}(x; \upsilon) = \max\{\langle \nabla U_{1}(x), \upsilon\rangle, \langle \nabla U_{2}(x), \upsilon\rangle\}$ for all $x\in \Gamma$.
\end{lemma}

\section{Problem statement} \label{sec: prob-statment}
Consider the following nonlinear time-invariant singularly perturbed system
\begin{align}
\dot{x} = f(x,z,u), \hspace{12pt} \epsilon\dot{z} = g(x,z,u), \label{non-linear}
\end{align}
where $ x \in \mathbb{R}^{n_{x}}$ and $z \in \mathbb{R}^{n_{z}}$ are the states, $u \in \mathbb{R}^{n_{u}}$ is the control input and $\epsilon > 0$ is a small parameter. We use singular perturbation theory to approximate the slow and the fast dynamics. We rely on the following standard assumption (see (11.3)-(11.4) in \cite{Khalil}).
\begin{ass} \label{ass: unique root}
The equation $g(x,z,u) = 0$ has $n\geq1$ isolated real roots
\begin{equation}\label{isolated-roots}
  z = h_{i}(x,u), \hspace{10pt} i = 1,2, ... , n,
\end{equation}
where $h_{i}$ is continuously differentiable.
\end{ass}
In that way, the substitution of the $i$th root $z = h(x,u)$ into the $x$-system yields the corresponding approximate slow model
\begin{equation} \label{xs}
\dot{x} = f(x,h(x,u),u).
\end{equation}
To investigate stability, it is more convenient to write system (\ref{non-linear}) with the coordinates $(x,y)$ where
\begin{equation}\label{non-linear-y}
y := z - h(x,u)
\end{equation}
is introduced to shift the quasi-steady-state of $z$ to the origin. Then we derive the approximate fast dynamics
\begin{align}\label{non-linear-fast1}
\frac{dy}{d\tau} = g(x,y+h(x,u),u),
\end{align}
where $\tau := (t-t_{0})/\epsilon$ and $x\in\mathbb{R}^{n_{x}}$ is treated as a fixed parameter.

In this study, we want to stabilize system (\ref{non-linear}) using a controller which is implemented over a network. We opt for an event-triggered implementation in the sense that transmissions are not triggered by a clock but according to a state-dependent criterion, which may reduce the utilization of the network compared to the periodic approach. Moreover, we concentrate on the case where the approximate fast dynamics (\ref{non-linear-fast1}) is stable and we aim at designing the feedback law based only on the slow model (\ref{xs}) as explained in Section \ref{itroduction}.

We follow an emulation-like approach as we first assume that the slow model (\ref{xs}) can be stabilized by a controller of the form $u=k(x)$. Afterwards, we take into account the effects of the network and we synthesize appropriate triggering conditions. The controller receives the state measurements only at the transmission instants $t_{i}, i\in\mathbb{Z}_{\geq 0}$ and we consider zero-order-hold devices. In that way $u(t) = k(x(t_{i}))$ for all $t \in [t_{i}, t_{i+1})$. The sequence of transmission instants $t_{i}, i\in\mathbb{Z}_{\geq 0}$ is defined by the event-triggering condition we will design. We introduce the sampling-induced error $e \in \mathbb{R}^{n_{x}}$ as in \cite{Tabuada2007event}, which is defined by $e(t) := x(t_{i}) - x(t)$ for all $t \in [t_{i}, t_{i+1})$ and which is reset to zero at each transmission instant. The state feedback controller is therefore given by
\begin{equation}\label{u}
u = k(x + e).
\end{equation}
Hence, in view of (\ref{non-linear-y}), the variable $y$ becomes
\begin{equation}\label{non-linear-y2}
y = z - h(x,k(x+e)).
\end{equation}
As a consequence, the system in the $(x,y)$ coordinates is, for all $t \in [t_{i}, t_{i+1}), i \in \mathbb{Z}_{\geq 0}$
\begin{align}
\dot{x}\!&=\!f\Big(x,y+h(x,k(x+e)),k(x+e)\Big)\!=:\!f_{x}(x,y,e) \label{x-system}\\
\epsilon\dot{y}\!&=\!g\Big(\!x, y\!+\!h(x,\!k(x\!+\!e)), k(x\!+\!e)\!\Big)\!-\!\epsilon\frac{\partial h}{\partial x}f_{x}(x,y,e) \nonumber\\
                &=: f_{y}(x,y,e), \label{fy}
\end{align}
and at each transmission
\begin{align}
x(t_{i+1}^{+}) &= x(t_{i+1}) \\
y(t_{i+1}^{+}) &= z(t_{i+1}^{+}) - h\Big(x(t_{i+1}^{+}),k(x(t_{i+1}^{+})+e(t_{i+1}^{+}))\Big) \nonumber \\
               &= z(t_{i+1}) - h\Big(x(t_{i+1}),k(x(t_{i+1}))\Big) \nonumber \\
               &= y(t_{i+1}) + h\Big(x(t_{i+1}),k(x(t_{i+1})+e(t_{i+1}))\Big) \nonumber\\
               &\quad - h\Big(x(t_{i+1}),k(x(t_{i+1}))\Big)  \nonumber\\
               &=: h_{y}(x(t_{i+1}),y(t_{i+1}),e(t_{i+1})). \label{y-jumps}
\end{align}
It has been shown in \cite{Romain2011unifying} that additional variables may be useful when designing the event-triggering condition. This will be the case for one of the strategies we propose in Section \ref{sec: main-results}. We denote these extra variables by a single vector $\tau\in \mathbb{R}^{n_{\tau}}$. The problem is modeled using the hybrid formalism of \cite{Teel} (like in \cite{Donkers2012output}, \cite{Romain2011unifying})
\begin{equation} \label{hybrid-model}
    \dot{q} = F(q) \hspace{0.2cm} \text{ for } q\in C, \hspace{0.6cm} q^{+} = G(q) \hspace{0.2cm} \text{ for } q\in D,
\end{equation}
where $q = (x,y,e,\tau) \in \mathbb{R}^{n_{q}}, n_{q} = 2n_{x} + n_{y} + n_{\tau}$,
\begin{equation} \label{flow_jump_map}
F(q) := \left(\begin{array}{c}
           f_{x}(x,y,e)\\
           \frac{1}{\epsilon}f_{y}(x,y,e) \\
           -f_{x}(x,y,e)\\
           f_{\tau}(x,y,e,\tau)
         \end{array}\right), \hspace{0.2cm}
G(q) := \left(\begin{array}{c}
           x\\
           h_{y}(x,y,e) \\
           0\\
           h_{\tau}(x,y,e,\tau)
         \end{array}\right),
\end{equation}
and $f_{\tau}$ and $h_{\tau}$ are designed vector fields which respectively define the dynamics of $\tau$ on flows and at jumps.
The flow set $C$ and the jump set $D$ in (\ref{hybrid-model}) are defined according to the event-triggering condition which we will synthesize in the following. The system flows on $C$ where the triggering condition is not satisfied and experiences a jump on $D$ where the triggering condition is verified. When $q \in C \cap D$, the system can either jump or flow, the latter only if flowing keeps $q$ in $C$. The flow map $F$ and the jump map $G$ are assumed to be continuous and the sets $C$ and $D$ will be closed (which ensures that system (\ref{hybrid-model}) is well-posed, see Chapter 6 in \cite{Teel}). We note that the state variable $y$ experiences a jump on the set $D$ according to (\ref{flow_jump_map}).

We briefly recall some basics about the hybrid formalism of \cite{Teel}. A set $E \subset \mathbb{R}_{\geq0}\times \mathbb{Z}_{\geq 0}$ is called a compact hybrid time domain if $E =  \displaystyle \underset{j\in\{0,...,J\}}{\cup}([t_{j}, t_{j+1}], j)$ for some finite sequence of times $0=t_{0}\leq t_{1} \leq ... \leq t_{J}$ and it is a hybrid time domain if for all $(T,J)\in E, E \cap ([0,T]\times \{0,1,...,J\})$ is a compact hybrid time domain. A function $\phi: E\rightarrow\mathbb{R}^{n_{q}}$ is a hybrid arc if $E$ is a hybrid time domain and if for each $j\in\mathbb{Z}_{\geq 0}, t \mapsto \phi(t,j)$ is locally absolutely continuous on $I^{j} := \{ t: (t,j)\in E \}$. A hybrid arc $\phi$ is a solution to (\ref{hybrid-model}) if: (i) $\phi(0,0)\in C\cup D$; (ii) for any $j\in \mathbb{Z}_{\geq 0}$, $\phi(t,j) \in C$ and $\dot{\phi}(t,j) = F(\phi(t,j))$ for almost all $t \in I^{j}$; (iii) for every $(t,j)\in \dom \phi$ such that $(t,j+1)\in \dom \phi$, it holds that $\phi(t,j) \in D$ and $\phi(t,j+1) = G(\phi(t,j))$. A solution $\phi$ to (\ref{hybrid-model}) is maximal if there does not exist another solution $\psi$ to (\ref{hybrid-model}) such that $\phi$ is a truncation of $\psi$ to some proper subset of $\dom \psi$.

Our objective is to design event-triggering conditions for system (\ref{hybrid-model}), which is equivalent to defining the sets $C$ and $D$, to guarantee stability properties for system (\ref{hybrid-model}) and the existence of a uniform amount of time between two jumps, which is essential in practice as the hardware cannot generate transmissions that are arbitrarily close in time. Moreover, we want to ignore the fast dynamics, which will be assumed to be stable, and design triggering conditions based only on the slow variables, i.e. the state $x$, and the sampling-induced error $e$ and potentially an additional designed variable $\tau$.

\section{Assumptions} \label{subsec: assumptions}
We present the assumptions made on system (\ref{hybrid-model}). We will show in Section \ref{sec: application-glob-Lips} that all the conditions are satisfied by a class of globally Lipschitz systems. The approximate slow and fast models (\ref{xs}) and (\ref{non-linear-fast1}) are now in view of (\ref{x-system}) and (\ref{fy})
\begin{align}
\dot{x}\!&=\!f\Big(x,h(x,k(x+e)),k(x+e)\Big)\!=:\!f_{s}(x,e) \label{non-linear-slow}\\
\frac{dy}{d\tau}\!&=\!g\Big(\!x,y\!+\!h(x,\!k(x\!+\!e)),k(x\!+\!e)\!\Big)\!=:\!g_{f}(x,y,e). \label{non-linear-fast}
\end{align}
First, we assume that the slow system (\ref{non-linear-slow}) is input-to-state stable (ISS) with respect to $e$.
\begin{ass}\label{ass: assumption-Lyapunov-Vx}
There exist a continuously differentiable function $V_{x}: \mathbb{R}^{n_{x}}\rightarrow \mathbb{R}_{ \geq 0}$ and class $\mathcal{K}_{\infty}$ functions $\underline{\alpha}_{x}, \overline{\alpha}_{x}, \gamma_{1}$ with $\gamma_{1}$ continuously differentiable and $\alpha_{1} > 0$ such that for all
$(x,e) \in \mathbb{R}^{2n_{x}}$ the following is satisfied
\begin{equation} \label{Vx-bounds}
\begin{array}{rcl}
\underline{\alpha}_{x}(|x|) & \leq & V_{x}(x) \leq \overline{\alpha}_{x}(|x|) \\
\frac{\partial V_{x}}{\partial x}f_{s}(x,e) & \leq & -\alpha_{1}V_{x}(x) + \gamma_{1}(|e|).
\end{array}
\end{equation}
\end{ass}

Since we design the triggering condition based only on the slow dynamics, to guarantee the overall stability of the closed-loop system, we need to make some assumptions on the stability of the \emph{approximate} fast model (\ref{non-linear-fast}) on flows. In particular, we assume that the following stability property holds for the \emph{approximate} fast dynamics like in \cite{Khalil}.
\begin{ass}\label{ass: assumption-Lyapunov-Vy}
There exist a continuously differentiable function $V_{y}: \mathbb{R}^{n_{y}} \rightarrow \mathbb{R}_{\geq 0}$ and class $\mathcal{K}_{\infty}$ functions $\underline{\alpha}_{y},
\overline{\alpha}_{y}$ and $\alpha_{2} > 0$ such that for all $(x,y,e) \in \mathbb{R}^{2n_{x} + n_{y}}$
\begin{equation} \label{Vy-bounds}
\begin{array}{rcl}
\underline{\alpha}_{y}(|y|) & \leq & V_{y}(x,y) \leq \overline{\alpha}_{y}(|y|) \\
\frac{\partial V_{y}}{\partial y}g_{f}(x,y,e) & \leq & -\alpha_{2}V_{y}(x,y).
\end{array}
\end{equation}
\end{ass}
Assumption \ref{ass: assumption-Lyapunov-Vy} implies that the origin of the \emph{approximate} fast dynamics (\ref{non-linear-fast}) is globally asymptotically stable. Note that Assumption \ref{ass: assumption-Lyapunov-Vy} does not imply that the origin of the fast dynamics (\ref{non-linear-fast}) is globally exponentially stable as the functions $\underline{\alpha}_{y}, \overline{\alpha}_{y}$ can be nonlinear. We impose the following conditions on the interconnections between the slow and fast dynamics (\ref{non-linear-slow}), (\ref{non-linear-fast}).
\begin{ass}\label{ass: assumption-Lyapunov-Interconnection-terms}
There exist a class $\mathcal{K}_{\infty}$ function $\gamma_{2}$ and $\beta_{1}, \beta_{2}, \beta_{3} > 0$ such that for all $(x,y,e) \in \mathbb{R}^{2n_{x} + n_{y}}$ the following hold
\begin{equation}\label{interconnection}
\begin{array}{c}
\frac{\partial V_{x}}{\partial x}\!\left[ f_{x}(x,y,e) -  f_{s}(x,e)\right] \leq \beta_{1}\sqrt{V_{x}(x)V_{y}(x,y)} \\
\left[\!\frac{\partial V_{y}}{\partial x}\!-\!\frac{\partial V_{y}}{\partial y}\frac{\partial h}{\partial x}\!\right]\!f_{x}(x,\!y,\!e) \leq \beta_{2}\sqrt{V_{x}(x)V_{y}(x,\!y)}\!+\!\beta_{3}V_{y}(x,\!y)\\
\qquad + \gamma_{2}(|e|),
\end{array}
\end{equation}
where $V_{x}$ and $V_{y}$ come from Assumptions \ref{ass: assumption-Lyapunov-Vx} and \ref{ass: assumption-Lyapunov-Vy} respectively.
In addition, there exists $L > 0$ such that, for all $s \geq 0$
\begin{equation}\label{cond-asm5}
\gamma_{2} \circ \gamma_{1}^{-1}(s) \leq Ls,
\end{equation}
where $\gamma_{1}$ comes from Assumption \ref{ass: assumption-Lyapunov-Vx}.
\end{ass}
\noindent Conditions (\ref{interconnection}) represent the effect of the deviation of the original system (\ref{hybrid-model}) from the slow and fast models (\ref{non-linear-slow}), (\ref{non-linear-fast}) respectively and are related to (11.43) and  (11.44) in \cite{Khalil}.

Finally, we assume that the dynamics of $V_{y}$ along jumps of system (\ref{hybrid-model}) satisfies the following condition.

\begin{ass}\label{ass: Vy-leq-lambdaV}
There exist $\lambda_{1},\lambda_{2} > 0$ such that for all $(x,y,e) \in \mathbb{R}^{2n_{x} + n_{y}}$
\begin{align}
V_{y}(x,h_{y}(x,y,e)) &\leq V_{y}(x,y) + \lambda_{1}\gamma_{1}(|e|) \nonumber\\
&\quad + \lambda_{2}\sqrt{\gamma_{1}(|e|)V_{y}(x,y)},
\end{align}
where $V_{x}, \gamma_{1}$ and $V_{y}$ come from Assumptions \ref{ass: assumption-Lyapunov-Vx} and \ref{ass: assumption-Lyapunov-Vy} respectively.
\end{ass}
Assumption \ref{ass: Vy-leq-lambdaV} is an algebraic condition which only requires the knowledge of $h_{y}$ (which is defined in (\ref{y-jumps})) and $\gamma_{1}$ and $V_{y}$ from Assumptions \ref{ass: assumption-Lyapunov-Vx} and \ref{ass: assumption-Lyapunov-Vy} respectively: we do not need to know the triggering condition to check it.

\begin{rmk}
Assumptions \ref{ass: assumption-Lyapunov-Vy}, \ref{ass: assumption-Lyapunov-Interconnection-terms} may require (\ref{Vy-bounds}), (\ref{interconnection}) to hold regardless the magnitude of the sampling-induced error $e$. We show in Section \ref{sec: application-glob-Lips} that all these conditions are satisfied by a class of globally Lipschitz systems which encompasses LTI systems as a particular case and for which these results are new.
\end{rmk}

\section{Main results} \label{sec: main-results}
First, we show that the design of triggering conditions of the same form as in \cite{Tabuada2007event} for the slow model may not ensure the existence of a strictly positive minimum amount of time between two jumps for the overall system. We then present our main results.

\subsection{A first observation}\label{subsec: first-obs}
In view of Assumption \ref{ass: assumption-Lyapunov-Vx}, a first attempt would be to define a triggering condition of the form $\gamma_{1}(|e|) \geq \sigma \alpha_{1}V_{x}(x)$ where $\sigma \in (0,1)$ like in \cite{Tabuada2007event}. The flow and jump sets are in this case
\begin{equation}
\begin{aligned}
C &= \{ q: \gamma_{1}(|e|) \leq \sigma \alpha_{1}V_{x}(x) \} \\
D &= \{ q: \gamma_{1}(|e|) = \sigma \alpha_{1}V_{x}(x) \}.
\end{aligned}
\end{equation}
The results in \cite{Tabuada2007event} guarantee the global asymptotic stability of the origin of the slow model (\ref{non-linear-slow}) and the existence of a uniform (semiglobal) amount of time between two jumps (under some conditions). However, this triggering rule no longer ensures a minimum time of flow between two jumps for system (\ref{hybrid-model}). Indeed $G(D) \cap D = \{q\,:\,x=e=0\}\neq \emptyset$. Thus, any solution in $G(D) \cap D$ may jump an infinite number of times, which makes the controller not realizable in practice. In the sequel, we first apply existing strategies in order to overcome this issue and we investigate how to modify the stability analysis and what kind of stability property one may expect. We also propose another strategy that allows to guarantee global asymptotic stability.

\subsection{Semiglobal practical stabilization}\label{subsec: practical-stability}
The most straightforward approach to enforce a lower bound on the inter-jumps for system (\ref{hybrid-model}) is to add a dead-zone to the triggering condition in Section \ref{subsec: first-obs}, i.e.
\begin{equation} \label{eq: trig-conditionPS}
\gamma_{1}(|e|) \geq \max \{\sigma \alpha_{1}V_{x}(x), \rho \},
\end{equation}
where $\rho > 0$ is a design parameter. The flow and jump sets in (\ref{hybrid-model}) are then
\begin{equation}\label{C-D-sets}
\begin{aligned}
C &= \{ q: \gamma_{1}(|e|) \leq \max \{\sigma \alpha_{1}V_{x}(x), \rho \} \} \\
D &= \{ q: \gamma_{1}(|e|) = \max \{\sigma \alpha_{1}V_{x}(x), \rho \} \}
\end{aligned}
\end{equation}
and we do not need to introduce an extra variable $\tau$, i.e. $q=(x,y,e)$.
Although this type of triggering conditions has already been used in \cite{Donkers2012output}, \cite{Mazo2012decentralized} for example, the fact that the state $y$ experiences jumps and that we rely on different assumptions require to fully modify the stability analysis and leads to the following result.

\begin{thm} \label{thm: practical-stability}
Consider system (\ref{hybrid-model}) with the flow and jump sets defined in (\ref{C-D-sets}). Suppose that Assumptions \ref{ass: unique root}-\ref{ass: Vy-leq-lambdaV} hold. Then, for any $\Delta, \rho > 0$, there exist $\beta \in \mathcal{KL}$, $\kappa \in \mathcal{K}_{\infty}$ and $\epsilon^{*}>0$ such that for any $\epsilon \in (0,\epsilon^{*})$ and any solution $\phi = (\phi_{x},\phi_{y},\phi_{e})$ with $|\phi(0,0)|\leq\Delta$,
\begin{align} \label{practical-stability}
|\phi(t,j)| &\leq \beta(|\phi(0,0)|,t+j) + \kappa(\rho) \hspace{10pt} \forall(t,j) \in \textnormal{dom } \phi,
\end{align}
and all inter-transmission times are lower-bounded by a strictly positive constant $\frac{\rho}{\xi(\Delta)}$, where $\xi: \mathbb{R}_{\geq 0} \rightarrow  \mathbb{R}_{>0}$ is a continuous increasing function, i.e. for all $j \in \mathbb{Z}_{\geq 0}$ $\sup I^{j} - \inf I^{j} \geq \frac{\rho}{\xi(\Delta)}$, where $I^{j} = \{ t: (t,j)\in \textnormal{dom } \phi \}$. Furthermore, all maximal solutions to (\ref{hybrid-model}) are complete.
\end{thm}

Theorem \ref{thm: practical-stability} ensures a semiglobal practical stability property for system (\ref{hybrid-model}). Indeed, given an arbitrary (large) ball of initial conditions centered at the origin and of radius $\Delta$ and any constant $\rho$, there exists $\epsilon$ sufficiently small such that solutions to (\ref{hybrid-model}), (\ref{C-D-sets}) converge towards a neighbourhood of the origin whose `size' can be rendered arbitrarily small by reducing $\rho$ (at the price of shorter inter-transmission intervals).


\subsection{Global asymptotic stabilization}\label{subsec: mixing-clock}
We propose another strategy to design the event-triggering condition to ensure a global asymptotic stability property under an extra assumption. The idea is to combine the event-triggered technique of \cite{Tabuada2007event} with the time-triggered results of \cite{Nesic2009explicit} such that we allow transmissions only after a fixed amount of time $T^{*}$ has elapsed since the last one.

We suppose that Assumptions \ref{ass: unique root}-\ref{ass: Vy-leq-lambdaV} are satisfied with $\gamma_{1}(s)=\bar{\gamma}_{1}s^{2}$ and $\gamma_{2}(s)=\bar{\gamma}_{2}s^{2}$ for some $\bar{\gamma}_{1},\bar{\gamma}_{2}\geq0$ and for $s\geq0$. We introduce an extra variable $\tau \in \mathbb{R}_{\geq0}$ to model the elapsed time between two successive jumps. Hence, $\dot{\tau} = 1$ on flows and $\tau^{+} = 0$ at jumps. Consequently, $q = (x,y,e,\tau)$ and $f_{\tau}(x,y,e,\tau) = 1$, $h_{\tau}(x,y,e,\tau) = 0$ in (\ref{flow_jump_map}). We define the flow and jump sets as follows
\begin{equation} \label{flow-jump-sets-clock}
\begin{array}{l}
C := \{q: \bar{\gamma}_{1}|e|^{2} \leq \sigma \alpha_{1}V_{x}(x) \text{ or } \tau \in [0,T^{*}]\} \\
D := \Big\{q: \Big(\bar{\gamma}_{1}|e|^{2} = \sigma \alpha_{1}V_{x}(x) \text{ and } \tau \geq T^{*}\Big) \text{ or } \\
\hspace{1.57cm} \Big(\bar{\gamma}_{1}|e|^{2} \geq \sigma \alpha_{1}V_{x}(x) \text{ and } \tau = T^{*}\Big)\Big\}.
\end{array}
\end{equation}
While the idea of merging event-triggered and time-triggered techniques is intuitive, the stability analysis is non-trivial as we need to build a common hybrid Lyapunov function for the two approaches. It has to be emphasized that the constant $T^{*}$ allows us to directly tune the minimum inter-transmission interval provided it is smaller than the bound given below.

Inspired by \cite{Nesic2009explicit}, we make the following additional assumption on system (\ref{hybrid-model}).
\begin{ass}\label{ass: clock}
There exist $M, N \geq 0$ such that, for all $(x,y)\in \mathbb{R}^{n_{x}+n_{y}}$ and for almost all $e \in \mathbb{R}^{n_{x}}$
\begin{equation*}
\langle \nabla|e|, -f_{x}(x,y,e)\rangle \leq M|e| + N(\sqrt{V_{x}(x)} + \sqrt{V_{y}(x,y)}),
\end{equation*}
where $V_{x}$ and $V_{y}$ come from Assumptions \ref{ass: assumption-Lyapunov-Vx} and \ref{ass: assumption-Lyapunov-Vy} respectively.
\end{ass}
\noindent The constant $T^{*}$ in (\ref{flow-jump-sets-clock}) is selected such that $T^{*} < \mathcal{T}$, like in \cite{Nesic2009explicit}, where
\begin{equation}\label{T-phi}
\mathcal{T} := \left\{
    \begin{array}{ll}
    \frac{1}{Mr}\arctan(r)& \hspace{10pt} M^{2} < \frac{\bar{\gamma}_{1}N^{2}}{\alpha_{1}} \\
    \frac{1}{M}& \hspace{10pt} M^{2} = \frac{\bar{\gamma}_{1}N^{2}}{\alpha_{1}} \\
    \frac{1}{Mr}\arctanh(r)& \hspace{10pt} M^{2} > \frac{\bar{\gamma}_{1}N^{2}}{\alpha_{1}}
    \end{array}
    \right.
\end{equation}
with $r := \sqrt{\left|\frac{\bar{\gamma}_{1}N^{2}}{\alpha_{1}M^{2}} - 1\right|}$, where $M, N$ come from Assumption \ref{ass: clock} and $\alpha_{1}, \bar{\gamma}_{1}$ come from Assumption \ref{ass: assumption-Lyapunov-Vx}. We obtain the following result.

\begin{thm} \label{thm: clock}
Consider system (\ref{hybrid-model}) with the flow and jump sets defined in (\ref{flow-jump-sets-clock}) and suppose the following hold.
\begin{enumerate}
  \item Assumptions \ref{ass: unique root}-\ref{ass: clock} hold with $\gamma_{1}(s)=\bar{\gamma}_{1}s^{2}$ and $\gamma_{2}(s)=\bar{\gamma}_{2}s^{2}$ with $\bar{\gamma}_{1},\bar{\gamma}_{2}\geq0$, for $s\geq0$.
  \item The constant $T^{*}$ in (\ref{flow-jump-sets-clock}) is such that $T^{*} \in (0,\mathcal{T})$.
\end{enumerate}
Then there exist $\beta \in \mathcal{KL}$ and $\bar{\epsilon}>0$ such that for any $\epsilon \in (0,\bar{\epsilon})$ and any solution $\phi = (\phi_{x},\phi_{y},\phi_{e},\phi_{\tau})$
\begin{equation} \label{eq-thm-clock}
|(\phi_{x}(t,j),\phi_{y}(t,j))| \leq \beta(|\phi(0,0)|,t+j) \hspace{10pt} \forall(t,j) \in \textnormal{dom } \phi.
\end{equation}
Moreover, all maximal solutions to (\ref{hybrid-model}) are complete.
\end{thm}
We see that Theorem \ref{thm: clock} ensures a global asymptotic stability property and that it requires an additional condition to hold, namely Assumption \ref{ass: clock}, compared to Section \ref{subsec: practical-stability}. 

\section{The case of globally Lipschitz systems} \label{sec: application-glob-Lips}
In this section, we show that all the conditions of Section \ref{subsec: assumptions} are verified by a class of globally Lipschitz systems, which includes LTI systems as a particular case. We assume that the vector fields $f, g$ and $k$ are globally Lipschitz and that the stability of the slow and the fast model can be verified using quadratic functions $V_{x}$ and $V_{y}$. Under these conditions, the proposition below states that Assumptions \ref{ass: unique root}-\ref{ass: clock} hold. Hence, the triggering rules presented in Sections \ref{subsec: practical-stability} and \ref{subsec: mixing-clock} can be applied.
\begin{props}\label{prop: glob-lip}
Consider system (\ref{non-linear}), (\ref{u}). Suppose the following hold
\begin{enumerate}
  \item The functions $f, g$ and $k$ are globally Lipschitz.
  \item Assumption \ref{ass: unique root} is verified with $h$ globally Lipschitz.
  \item There exist positive definite and symmetric real matrices $P_{1}, P_{2}$ such that the functions $V_{x}: x \mapsto x^{T}P_{1}x$ and $V_{y}: y \mapsto y^{T}P_{2}y$ satisfy for all $(x,y,e)\in \mathbb{R}^{2n_{x} + n_{y}}$
\end{enumerate}
\begin{gather}
    \frac{\partial V_{x}}{\partial x}f_{s}(x,0) \leq -\bar{\alpha}_{1}V_{x}(x) \label{eq: prop1-Vx-dot}\\
    \frac{\partial V_{y}}{\partial y}g_{f}(x,y,e) \leq -\alpha_{2}V_{y}(x,y),
  \end{gather}
\indent\indent where $\bar{\alpha}_{1}, \alpha_{2} > 0$. Then, Assumptions \ref{ass: assumption-Lyapunov-Vx}-\ref{ass: clock} are satisfied.
\end{props}
The choice of the triggering condition between Section \ref{subsec: practical-stability} and \ref{subsec: mixing-clock} has to be done on a case-by-case basis according to the desired specifications. The technique in Section \ref{subsec: mixing-clock} offers stronger a priori guarantees on the minimum inter-event times, however the strategy in Section \ref{subsec: practical-stability} may lead to less transmissions in average, see for simulation results on the autopilot event-triggered control of an F-8 aircraft \cite{Abdelrahim2013Event}.

\section{Conclusion}
We have investigated the event-triggered stabilization of nonlinear singularly perturbed systems based only on the slow dynamics. Two classes of controllers have been developed which ensure different asymptotic stability properties. We believe that this work can be extended along two important directions. First, the design of event-triggered controllers for singularly perturbed systems with potentially unstable fast dynamics can be pursued based on the model (\ref{hybrid-model}). Second, it would be interesting to use similar ideas as in this paper for the event-triggered control of systems based on a simplified model of the plant dynamics obtained by other means like model reduction.

\bibliographystyle{plain}
\bibliography{References}
%
%
\section*{Appendix}

\textbf{Proof of Theorem \ref{thm: practical-stability}}.
We define the function (like in the proof of Theorem 1 in \cite{Sanfelice2011On})
\begin{equation}\label{Lyapunov-overall}
 V(q) := V_{x}(x) + \sqrt{\epsilon}V_{y}(x,y) \hspace{0.5cm}\forall q\in \mathbb{R}^{n_{q}}
\end{equation}
with $\epsilon\in(0,\epsilon^{*})$ where $\epsilon^{*}>0$ will be defined in the following. Let $q\in C$,
it holds that, in view of (\ref{x-system}) and (\ref{fy}) (we omit the arguments of $f_{x}, g, V_{x}, V_{y}$ in the following for space reasons)
\begin{equation}
\begin{array}{lllllll}
\langle \nabla V(q), F(q) \rangle  & = & \frac{\partial V_{x}}{\partial x}f_{x} + \sqrt{\epsilon} \frac{\partial V_{y}}{\partial x}f_{x} + \frac{\sqrt{\epsilon}}{\epsilon}\frac{\partial V_{y}}{\partial y}f_{y} \\
        & = & \frac{\partial V_{x}}{\partial x}f_{s} + \frac{\partial V_{x}}{\partial x} \left[ f_{x} - f_{x_{s}} \right] + \sqrt{\epsilon}^{-1}\frac{\partial V_{y}}{\partial y}g_{f}  \\
        & &  + \sqrt{\epsilon} \left[ \frac{\partial V_{y}}{\partial x} - \frac{\partial V_{y}}{\partial y}\frac{\partial h}{\partial x} \right]f_{x}.
\end{array}
\end{equation}
In view of the definition of the set $C$, $\gamma_{1}(|e|) \leq \max\{\sigma\alpha_{1}V_{x}(x),\rho\}$ and $\gamma_{2}(|e|)\leq \gamma_{2}\circ\gamma_{1}^{-1}\big(\max\{\sigma\alpha_{1}V_{x}(x),$ $\rho\}\big)$. The condition (\ref{cond-asm5}) ensures that $\gamma_{2}(|e|)\leq L\max\{\sigma\alpha_{1}V_{x}(x),\rho\}$. Using Assumptions \ref{ass: assumption-Lyapunov-Vx}-\ref{ass: Vy-leq-lambdaV}, we derive that
\begin{equation}
\begin{array}{lllllll}
\langle \nabla V(q), F(q) \rangle  & \leq & -\chi^{T}\mathcal{A}\chi  + (1+\sqrt{\epsilon}L)\rho
\end{array}
\end{equation}
where $\chi:=(\sqrt{V_{x}(x)},\,\sqrt{V_{y}(x,y)})$ and
\begin{equation}
\begin{array}{lllllll}
\mathcal{A}  & := & \left[
           \begin{array}{cc}
            \alpha_{1}(1-\sigma(1+\sqrt{\epsilon}L))  & -(\beta_{1}+\sqrt{\epsilon}\beta_{2})/2 \\
            -(\beta_{1}+\sqrt{\epsilon}\beta_{2})/2 & \alpha_{2}\sqrt{\epsilon}^{-1} -\sqrt{\epsilon}\beta_{3} \\
           \end{array}
         \right].
\end{array}
\end{equation}
Let $\mu\in(0,\alpha_{1}(1-\sigma))$. The following conditions ensure that $\mathcal{A}\geq \mu\text{diag}(1,\sqrt{\epsilon})$, where $\text{diag}(1,\sqrt{\epsilon})$ is the diagonal matrix with elements $(1,\sqrt{\epsilon})$ on the diagonal
\begin{equation}\label{eq: A-diag-mu}
\left\{
\begin{array}{l}
\alpha_{1}(1-\sigma(1+\sqrt{\epsilon}L)) \geq \mu\\
\Big(\alpha_{1}(1-\sigma(1+\sqrt{\epsilon}L))- \mu\Big)\Big(\alpha_{2}\sqrt{\epsilon}^{-1} \!-\!\sqrt{\epsilon}\beta_{3}- \sqrt{\epsilon}\mu\Big)\\
 \hfill \!\geq\!  (\beta_{1}\!+\!\sqrt{\epsilon}\beta_{2})^2/4
\end{array}\right.
\end{equation}
The inequalities in (\ref{eq: A-diag-mu}) are always satisfied by selecting $\epsilon^{*}>0$ (and so $\epsilon$) sufficiently small. Consequently
\begin{equation}
\begin{array}{lllllll}
\langle \nabla V(q), F(q) \rangle  & \leq & -\mu\chi^{T}\text{diag}(1,\sqrt{\epsilon})\chi  + (1+\sqrt{\epsilon}L)\rho\\
& = & -\mu V(q) + (1+\sqrt{\epsilon}L)\rho,
\end{array}
\end{equation}
from which we deduce that $V(q) \geq 2\frac{(1+\sqrt{\epsilon}L)}{\mu}\rho$ implies
\begin{equation}
\begin{array}{lllll}
\langle \nabla V(q), F(q) \rangle & \leq & -\frac{\mu}{2} V(q).
\end{array}\label{eq: thm1-V-flow}
\end{equation}
Let $q\in D$, $V(G(q)) = V_{x}(x) + \sqrt{\epsilon} V_{y}(x,h_{y}(x,y,e))$. In view of Assumption \ref{ass: Vy-leq-lambdaV} and the definition of the set $D$, $V_{y}(x,h_{y}(x,y,e)) \leq V_{y}(x,y) + \lambda_{1}\gamma_{1}(|e|) + \lambda_{2}\sqrt{\gamma_{1}(|e|)V_{y}(x,y)} = V_{y}(x,y) + \lambda_{1}\max\{\sigma \alpha_{1} V_{x}(x), \rho\} + \lambda_{2}\sqrt{\max\{\sigma \alpha_{1} V_{x}(x), \rho\}V_{y}(x,y)} \leq V_{y}(x,y) + \lambda_{1}\sigma \alpha_{1} V_{x}(x) +  \lambda_{1}\rho + \lambda_{2}\sqrt{(\sigma \alpha_{1} V_{x}(x)+\rho)V_{y}(x,y)}$. Using that\\ $\sqrt{\max\{\sigma \alpha_{1} V_{x}(x),\rho\}V_{y}(x,y)}\leq
\epsilon^{-\frac{1}{4}}\max\{\sigma \alpha_{1} V_{x}(x),\rho\}+\epsilon^{\frac{1}{4}}V_{y}(x,y)$, we deduce that
$V_{y}(x,h_{y}(x,y,e)) \leq V_{y}(x,y) + \lambda_{1} \sigma \alpha_{1}V_{x}(x) + \lambda_{1}\rho + \lambda_{2}\epsilon^{-\frac{1}{4}}\max\{\sigma \alpha_{1} V_{x}(x),\rho\}  + \lambda_{2}\epsilon^{\frac{1}{4}}V_{y}(x,y)$.
As a consequence, by taking $\epsilon^{*}$ sufficiently small: $\epsilon^{\frac{1}{2}}\leq\epsilon^{\frac{1}{4}}$ and $V(G(q)) \leq V(q) + \epsilon^{\frac{1}{4}}\lambda_{1} \sigma \alpha_{1}V_{x}(x) + \epsilon^{\frac{1}{4}}\lambda_{1}\rho + \epsilon^{\frac{1}{4}}\lambda_{2}\max\{\sigma \alpha_{1} V_{x}(x),\rho\}  + \epsilon^{\frac{1}{4}}\lambda_{2}\sqrt{\epsilon}V_{y}(x,y)$. After some direct computations, we derive that $V(G(q))\leq (1+\epsilon^{\frac{1}{4}}\lambda)V(q)+\epsilon^{\frac{1}{4}}\lambda\rho$ where $\lambda := (\lambda_{1}+\lambda_{2})\max\{\sigma\alpha_{1},1\}$. As a consequence
\begin{equation}
\begin{array}{lllll}
V(G(q)) & \leq & (1+2\epsilon^{\frac{1}{4}}\lambda)\max\{V(q),\rho\}.
\end{array}\label{eq: thm1-V-jump}
\end{equation}

Let $\Delta>0$ and $\phi = (\phi_{x}, \phi_{y}, \phi_{e})$ be a solution to (\ref{hybrid-model}), (\ref{C-D-sets}) such that $|\phi(0,0)|\leq \Delta$. Assume without loss of generality\footnote{If that is not the case, the inequality obtained later in (\ref{eq: thm1-V-phi}) will hold for any $(t,j)\in\dom \phi$ with $j\geq 1$. A bound on $V(\phi)$ on the interval $[0,t_{1}]$ can then be derived using (\ref{eq: thm1-V-flow}) and (\ref{eq: thm1-V-jump}) to upper-bound on $V(\phi)$ on the whole domain $\dom \phi$. Note that if $\phi$ never jumps, the bound on the inter-jump times used in (\ref{eq: thm1-cond-tau}) trivially holds and (\ref{eq: thm1-V-phi}) will be verified.} that $\phi_{e}(0,0)=0$.
By invoking standard comparison principles, we obtain from (\ref{eq: thm1-V-flow}), for all $(t,0)\in\dom \phi$,
\begin{equation}
\begin{array}{lllll}
V(\phi(t,0)) & \leq &  \max\big\{e^{-\frac{\mu}{2}t}V(\phi(0,0)),2\frac{(1+\sqrt{\epsilon}L)}{\mu}\rho\}\\
& \leq & \max\big\{V(\phi(0,0)),\theta\rho\}
\end{array}\label{eq: thm1-bound-t-0}
\end{equation}
with $\theta:=(1+2\lambda)\max\{2\frac{(1+L)}{\mu},1\big\}$ (where we have used the fact $\epsilon^{*}$ is sufficiently small such that $\epsilon^{*}\leq 1$). The length of the inter-jump interval is lower bounded by the time it takes for $\gamma_{1}(|\phi_{e}|)$ to grow from $0$ to $\rho$ in view of (\ref{C-D-sets}). We aim at establishing a lower bound of the latter time, which we denote $\tau(\Delta)$. In view of (\ref{eq: thm1-bound-t-0}), $(\phi_{x}(t,0),\phi_{y}(t,0))$ lie in the compact set $\{(x,y)\,:\,V(x,y,0)\leq \max\big\{\overline{\alpha}(\Delta),\theta\rho\}\}$ for all $(t,0)\in\dom \phi$,
where $\overline{\alpha}(|(\phi_{x}(t,j), \phi_{y}(t,j))|)=\overline{\alpha}_{x}(|\phi_{x}(t,j)|)+\overline{\alpha}_{y}(|\phi_{y}(t,j)|)$ (we use again the fact that $\epsilon^{*}\leq 1$ as well as (\ref{Vx-bounds}) and (\ref{Vy-bounds})). As $\phi_{e}(t,0)=\phi_{x}(0,0)-\phi_{x}(t,0)$, we deduce that $\phi(t,0)$ lie in a compact set $\mathcal{S}(\Delta)$ for all $(t,0)\in\dom\phi$.
Since $\gamma_{1}$ is continuously differentiable by assumption, $\phi$ is continuous between two jump instants, $f_{x}$ is continuous and $\mathcal{S}(\Delta)$ is compact
\begin{equation}
\begin{array}{lllll}
\frac{d}{dt}\gamma_{1}(|\phi_{e}(t,0)|) & \leq & \partial\gamma_{1}(|\phi_{e}(t,0)|)|f_{x}(\phi(t,0))| \\
                                & \leq & \displaystyle \sup_{\tiny\substack{q \in \mathcal{S}(\Delta)}}\left\{\partial\gamma_{1}(|e|)|f_{x}(x,y,e)|\right\}\\
                                & < & \xi(\Delta),
\end{array}
\end{equation}
for some $\xi(\Delta)>0$, which ensures the property on the inter-jump intervals stated below (\ref{practical-stability}). Hence $\tau(\Delta)\geq \frac{\rho}{\xi(\Delta)}$. To compensate the potential increase of $V$ at jumps in view of (\ref{eq: thm1-V-jump}), we will use the following condition
\begin{equation}
\begin{array}{lllll}
\tau(\Delta) & > & \frac{4}{\mu}\ln\big(1+2\epsilon^{\frac{1}{4}}\lambda\big)
\end{array}\label{eq: thm1-cond-tau}
\end{equation}
which is always satisfied by selecting $\epsilon^{*}$ sufficiently small such that $\frac{4}{\mu}\ln\big(1+2(\epsilon^{*})^{\frac{1}{4}}\lambda\big)\leq \frac{\rho}{\xi(\Delta)} \leq \tau(\Delta)$. The inequality (\ref{eq: thm1-cond-tau}) ensures $e^{-\frac{\mu}{2}t_{1}}(1+2\epsilon^{\frac{1}{4}}\lambda)<1$ where $t_{1}$ is such that $(t_{1},0),(t_{1},1)\in\dom \phi$ (and verifies $t_{1}\geq \tau(\Delta)$). As a consequence, from (\ref{eq: thm1-V-jump}) and (\ref{eq: thm1-bound-t-0}), we deduce that
\begin{equation}\label{eq: thm1-bound-t-1}
\begin{array}{lllll}
V(\phi(t_{1},1)) & \leq &  (1+2\epsilon^{\frac{1}{4}}\lambda)\max\{V(\phi(t_{1},0)),\rho\}\\
& \leq & (1+2\epsilon^{\frac{1}{4}}\lambda)\max\big\{e^{-\frac{\mu}{2}t_{1}}V(\phi(0,0)),\\
& & \hspace{2.6cm} 2\frac{(1+\sqrt{\epsilon}L)}{\mu}\rho,\rho\big\} \\
& \leq & \max\big\{V(\phi(0,0)),\theta\rho\}
\end{array}
\end{equation}
which is the same right hand-side as in (\ref{eq: thm1-bound-t-0}). Hence, we apply the same arguments as above to derive that $t_{2}-t_{1}\geq \tau(\Delta)$, where $(t_{2},1),(t_{2},2)\in\dom \phi$. Consequently, by induction $t\geq \tau(\Delta)j$ for any $(t,j)\in\dom \phi$.
By using the comparison principle and the fact that $(1+2\epsilon^{\frac{1}{4}}\lambda)e^{-\frac{\mu}{2}\tau(\Delta)}\leq 1$, we obtain in view of (\ref{eq: thm1-V-flow})  and (\ref{eq: thm1-V-jump}), for any $(t,j)\in\dom \phi$
\begin{equation}
\begin{array}{lllll}
V(\phi(t,j)) & \leq & \max\big\{e^{-\frac{\mu}{2}t}(1+2\lambda\epsilon^{\frac{1}{4}})^{j}V(\phi(0,0)),\theta\rho\}.
\end{array}\label{eq: thm1-V-phi}
\end{equation}
The condition (\ref{eq: thm1-cond-tau}) ensures, since $t\geq \tau(\Delta)j$ for any $(t,j)\in\dom \phi$, that $e^{-\frac{\mu}{2}t}(1+2\epsilon^{\frac{1}{4}}\lambda)^{j} \leq e^{-\gamma(t+j)}$ with $\gamma \in \left(0, \frac{\frac{\mu}{2} - \ln(1+2\epsilon^{\frac{1}{4}}\lambda)\frac{1}{\tau(\Delta)}}{1 + \frac{1}{\tau(\Delta)}}\right)$. As a consequence, for any $(t,j)\in\dom \phi$,
\begin{equation}
\begin{array}{lllll}
V(\phi(t,j)) & \leq & \max\big\{e^{-\gamma(t+j)}V(\phi(0,0)),\theta\rho\}.
\end{array}
\end{equation}
In view of (\ref{Vx-bounds}) and (\ref{Vy-bounds}), for any $(t,j)\in\dom \phi$,
\begin{equation}
\begin{array}{rllll}
\underline{\alpha}_{x}(|\phi_{x}(t,j)|) & \leq & \max\big\{e^{-\gamma(t+j)}\overline{\alpha}(\phi(0,0)),\theta\rho\}\\
|\phi_{x}(t,j)| & \leq & \underline{\alpha}_{x}^{-1}\left(\max\big\{e^{-\gamma(t+j)}\overline{\alpha}(|\phi(0,0)|),\theta\rho\}\right),
\end{array} \label{eq: thm1-phi_x}
\end{equation}
Using that $\gamma_{1}(|e|)\leq \max\{\sigma\alpha_{1}V(x),\rho\}$ for any $q\in C\cup D\cup G(D)$, we deduce that for any $(t,j)\in\dom \phi$
\begin{equation}
\begin{array}{rllll}
|\phi_{e}(t,j)| & \leq & \max\big\{\beta_{e}(|\phi(0,0)|,t+j),\vartheta_{e}(\rho)\}
\end{array}\label{eq: thm1-phi_e}
\end{equation}
for some $\beta_{e}\in \mathcal{KL}$ and $\theta_{e}\in \mathcal{K}_{\infty}$. We are left with the $y$-component of $\phi$. In view of Assumptions \ref{ass: assumption-Lyapunov-Vy}-\ref{ass: assumption-Lyapunov-Interconnection-terms}, it holds that
\begin{equation}
\begin{array}{lllll}
\langle \nabla V_{y}(x,y), (f_{x}, f_{y}) \rangle &=& \frac{1}{\epsilon}\frac{\partial V_{y}}{\partial y}g + \left[ \frac{\partial V_{y}}{\partial x} - \frac{\partial V_{y}}{\partial y}\frac{\partial h}{\partial x} \right]f_{x} \\
& &\hspace{-1cm} \leq -\frac{\alpha_{2}}{\epsilon}V_{y}(x,y) + \beta_{2}\sqrt{V_{x}(x)V_{y}(x, y)} \\
& &\hspace{-0.7cm} + \beta_{3}V_{y}(x,y) + \gamma_{2}(|e|) \\
& &\hspace{-1cm} \leq \!-(\frac{\alpha_{2}}{\epsilon}\!-\!\beta_{2}\!-\!\beta_{3})V_{y}(x,y) + \beta_{2}V_{x}(x) \\
& &\hspace{-0.7cm} + \gamma_{2}(|e|)
\end{array} \label{eq: thm1-Vy-flow}
\end{equation}
and, as shown before,
\begin{equation}
\begin{array}{lllll}
V_{y}(x,\!h_{y}\!(x,\!y,\!e)) &\!\leq\!&V_{y}\!(x,\!y)\!+\!\lambda_{1}\!\gamma_{1}(|e|)\!+\!\lambda_{2}\sqrt{\!\gamma_{1}(|e|)\!V_{y}(x,\!y)} \\
 &  & \hspace{-0.9cm} \leq (1 + \epsilon^{\frac{1}{4}}\lambda_{2})V_{y}(x,y) + (\lambda_{1} + \epsilon^{-\frac{1}{4}}\lambda_{2})\gamma_{1}(|e|)
\end{array}\label{eq: thm1-Vy-jump}
\end{equation}
By following similar lines as above, we deduce that, by taking $\epsilon^{*}$ sufficiently small, the $y$-system is ISS with respect to $x$ and $e$. As a consequence, in view of (\ref{eq: thm1-phi_x}), (\ref{eq: thm1-Vy-flow}) and (\ref{eq: thm1-Vy-jump}) we derive that
\begin{equation}
\begin{array}{rllll}
|\phi_{y}(t,j)| & \leq & \max\big\{\beta_{y}(|\phi(0,0)|,t+j),\vartheta_{y}(\rho)\}
\end{array}\label{eq: thm1-phi_y}
\end{equation}
for some $\beta_{y}\in \mathcal{KL}$ and $\theta_{y}\in \mathcal{K}_{\infty}$. The property (\ref{practical-stability}) then follows from (\ref{eq: thm1-phi_x}), (\ref{eq: thm1-phi_e}) and (\ref{eq: thm1-phi_y}). Equations (\ref{eq: thm1-phi_x}), (\ref{eq: thm1-phi_e}) and (\ref{eq: thm1-phi_y}) ensure that $\phi$ cannot explode in finite time, neither it can flow out of $C\cup D$ since $G(D)\subset C$. Noting that system (\ref{hybrid-model}), (\ref{C-D-sets}) does not admit trivial solution\footnote{This comes from the fact that $C\backslash D$ is the interior of $C$. Hence, the tangent cone (see definition 5.12 in \cite{Teel}) is $\mathbb{R}^{n}$ and (VC) in Proposition 6.10 in \cite{Teel} holds for any point in $C\backslash D$}, we conclude that maximal solutions to (\ref{hybrid-model}), (\ref{C-D-sets}) are complete according to Proposition 6.10 in \cite{Teel}. \hfill$\Box$


\textbf{Proof of Theorem \ref{thm: clock}}.
Like in the proof of Theorem 1 in \cite{Nesic2009explicit}, we introduce the solution $\zeta$ to
\begin{equation}
\begin{array}{lll}\label{zeta-dot}
\dot{\zeta} & = & -1 - 2M\zeta - \mu - \left(\mu\zeta(\tau)+\frac{\bar{\gamma}_{1}}{\alpha_{1}-\mu}(N \zeta)^2\right)\\
& =: & f_{\zeta}(\tau)
\end{array}
\end{equation}
with $\zeta(0) = \vartheta^{-1}$, $\vartheta\in(0,1)$ and $\mu\in(0,\alpha_{1})$. We denote $\widetilde{\mathcal{T}}(\mu,\vartheta)$ the time it takes for $\zeta$ to decrease from $\vartheta^{-1}$ to $\vartheta$. This time $\widetilde{\mathcal{T}}(\mu,\vartheta)$ is a continuous function of $\mu,\vartheta$ which is decreasing in $\mu, \vartheta$ (by invoking the comparison principle). On the other hand, we note that $\widetilde{\mathcal{T}}(\mu,\vartheta) \rightarrow \mathcal{T}$ as $(\mu, \vartheta)$ tends to $(0,0)$ by following similar lines as in the proof of Claim 1 in \cite{Nesic2009explicit}, where $\mathcal{T}$ is defined in (\ref{T-phi}). As a consequence, since $T^{*}< \mathcal{T}$, there exist $\mu,\vartheta$ such that $T^{*}\leq \widetilde{\mathcal{T}}(\mu,\vartheta)$ which we fix.

We define
\begin{equation} \label{R}
R(q) := V_{x}(x) + dV_{y}(x,y) + \max\{0,\bar{\gamma}_{1}\zeta(\tau)|e|^{2}\} \hspace{0.2cm}\forall q\in \mathbb{R}^{n_{q}}
\end{equation}
where $d\in\left(0,\min\{\frac{\bar{\gamma}_{1}}{\bar{\gamma}_{2}}\mu,\frac{1-\sigma}{\sigma}\frac{\bar{\gamma}_{1}}{\bar{\gamma}_{2}},\frac{\vartheta^{2}}{(\lambda_{1} + \lambda_{2})^{2}}, \frac{(e^{\mu T^{*}} - 1)^{2}}{\lambda^{2}},1\}\right)$ and $\lambda := \max\{\lambda_{2}, (\lambda_{1}+\lambda_{2})\sigma\alpha_{1}\}$.
Let $q\in C$ and consider the case where $\zeta(\tau) > 0$. In view of Assumptions \ref{ass: assumption-Lyapunov-Vx}-\ref{ass: clock} and Lemma \ref{lma: clarke}
\begin{equation}
\begin{array}{lllllll}
R^{\circ}(q;F(q)) & \leq & -(\chi,\,|e|)^{T}\mathcal{A}_{1}(\chi,\,|e|),
\end{array}
\end{equation}
where $\chi:=(\sqrt{V_{x}(x)},\,\sqrt{V_{y}(x,y)})$,\\[4pt] $\mathcal{A}_{1} := \left[\begin{smallmatrix}\alpha_{1} & -\frac{1}{2}(\beta_{1} + d\beta_{2})  & -\bar{\gamma}_{1} N \zeta(\tau)\\ \ast & \frac{d}{\epsilon}\alpha_{2} - d\beta_{3} & -\bar{\gamma}_{1} N \zeta(\tau)\\ \ast & \ast & \upsilon(\tau)           \end{smallmatrix} \right]$, $\upsilon(\tau):=-\bar{\gamma}_{1}-d\bar{\gamma}_{2}-\bar{\gamma}_{1} f_{\zeta}(\tau)-2\bar{\gamma}_{1} M \zeta(\tau)$ and $\ast$ stands for the symmetric components of $\mathcal{A}_{1}$.
The following conditions ensure that $\mathcal{A}_{1}\geq \mu\text{diag}(1,d,\bar{\gamma}_{1}\zeta(\tau))$ according to Sylvester's criterion, where $\text{diag}(1,d,\bar{\gamma}_{1}\zeta(\tau))$ is the diagonal matrix with elements $(1,d,\bar{\gamma}_{1}\zeta(\tau))$ on the diagonal
{\small\begin{equation}
\left\{\begin{array}{lll}
0 & \leq & \alpha_{1} - \mu  \\
0 & \leq & (\alpha_{1}-\mu)d(\frac{1}{\epsilon}\alpha_{2} - \beta_{3}-\mu) \geq \frac{1}{4}(\beta_{1} + d\beta_{2})^{2}\\
0 & \leq & (\alpha_{1}-\mu) \\
& & \hspace{-0.22cm}\times\left\{ d(\frac{1}{\epsilon}\alpha_{2}-\beta_{3}-\mu)(\upsilon(\tau)-\mu\bar{\gamma}_{1}\zeta(\tau)) - (\bar{\gamma}_{1}\zeta(\tau)N)^2 \right\}\\
& & \hspace{-0.22cm} +\frac{1}{2}(\beta_{1} + d\beta_{2}) \\
& & \hspace{-0.22cm} \times\left\{-\frac{1}{2}(\beta_{1}+d\beta_{2})(\upsilon(\tau)-\mu\bar{\gamma}_{1}\zeta(\tau))- (\bar{\gamma}_{1}\zeta(\tau)N)^2\right\}\\
& & \hspace{-0.22cm} -\bar{\gamma}_{1} N\zeta(\tau) \\
& & \hspace{-0.22cm} \times\left\{\frac{1}{2}(\beta_{1}+d\beta_{2})\bar{\gamma}_{1}\zeta(\tau)N + \bar{\gamma}_{1} N\zeta(\tau)d(\frac{1}{\epsilon}\alpha_{2} - \beta_{3}-\mu)\right\}.
\end{array}\right.
\end{equation}}
\hspace{-0.22cm}
The first two inequalities above are respectively verified by definition of $\mu$ and by taking $\epsilon$ sufficiently small. For the last inequality to hold, it suffices to select $\epsilon$ sufficiently small provided that $\frac{d}{\epsilon}\alpha_{2}\left((\alpha_{1}-\mu)(\upsilon-\mu\bar{\gamma}_{1}\zeta(\tau))-(\bar{\gamma}_{1} N\zeta(\tau))^2\right) > 0$ which is equivalent to, by definition of $\upsilon$ and definition of $f_{\zeta}$ in (\ref{zeta-dot}), $(\alpha_{1}-\mu)(\bar{\gamma}_{1}\mu-d\bar{\gamma}_{2}) > 0$ which holds by definition of $d$ and $\mu$. Consequently, by selecting $\epsilon$ sufficiently small
\begin{equation}
\begin{array}{lll}
R^{\circ}(q;F(q)) & \leq & -\mu R(q).
\end{array}\label{eq: thm2-R-flow}
\end{equation}
Suppose now that $\zeta(\tau) < 0$, hence $\bar{\gamma}_{1}|e|^{2}\leq\sigma\alpha_{1}V_{x}(x)$ in view of the definition of the set $C$. Using Assumptions \ref{ass: assumption-Lyapunov-Vx}-\ref{ass: clock} and Lemma \ref{lma: clarke},
\begin{equation}
\begin{array}{lllllll}
R^{\circ}(q;F(q)) & \leq & -\chi^{T}\mathcal{A}_{2}\chi,
\end{array}
\end{equation}
where $\mathcal{A}_{2} := \left[\begin{smallmatrix}\alpha_{1}\left(1-\sigma(1+d\bar{\gamma}_{2}\bar{\gamma}_{1}^{-1})\right)  & -\frac{1}{2}(\beta_{1} + d\beta_{2})  \\ -\frac{1}{2}(\beta_{1} + d\beta_{2}) & \frac{d}{\epsilon}\alpha_{2} - d\beta_{3} \end{smallmatrix}\right]$. By following similar arguments as above and since $d<\frac{1-\sigma}{\sigma}\frac{\bar{\gamma}_{1}}{\bar{\gamma}_{2}}$ and $R(q)=V_{x}(x)+dV_{y}(x,y)$ in this case, we derive that (\ref{eq: thm2-R-flow}) holds by selecting $\epsilon$ sufficiently small.
When $\zeta(\tau) = 0$, (\ref{eq: thm2-R-flow}) is verified in view of Lemma \ref{lma: clarke} and the results obtained for the cases where $\zeta(\tau)>0$ and $\zeta(\tau)<0$.

Let $q\in D$. Suppose that $\tau = T^{*}$ (note that $\bar{\gamma}_{1}|e|^{2} \geq \sigma \alpha_{1}V_{x}(x)$ in this case). In view of Assumption \ref{ass: Vy-leq-lambdaV} $R(G(q)) = V_{x}(x) + d V_{y}(x,h_{y}(x,y,e))\leq V_{x}(x) + d\Big(V_{y}(x,y) + \lambda_{1}\bar{\gamma}_{1}|e|^{2}+ \lambda_{2}\sqrt{\bar{\gamma}_{1}|e|^{2}V_{y}(x,y)}\Big)$. Using that $\sqrt{\bar{\gamma}_{1}|e|^{2}V_{y}(x,y)} \leq \frac{1}{\sqrt{d}}\bar{\gamma}_{1}|e|^{2} + \sqrt{d}V_{y}(x,y)$ and since $d \leq \sqrt{d} \leq 1$, it holds that $R(G(q)) \leq V_{x}(x) + dV_{y}(x,y) + \sqrt{d}(\lambda_{1} + \lambda_{2})\bar{\gamma}_{1}|e|^{2} + d\lambda_{2}\sqrt{d}V_{y}(x,y)\leq V_{x}(x) + dV_{y}(x,y) + \sqrt{d}(\lambda_{1} + \lambda_{2})\bar{\gamma}_{1}|e|^{2} + \lambda_{2}\sqrt{d}(V_{x}(x) + dV_{y}(x,y))$.
We take $d$ sufficiently small such that $\sqrt{d}(\lambda_{1} + \lambda_{2})\bar{\gamma}_{1}|e|^{2} \leq \bar{\gamma}_{1}\zeta(\widetilde{\mathcal{T}}(\mu,\vartheta))|e|^{2} = \bar{\gamma}_{1}\vartheta|e|^{2}$ (since $\zeta(\widetilde{\mathcal{T}}(\mu,\!\vartheta))\!$ $\!=\!\vartheta$). As a consequence $R(G(q)) \leq (1 + \lambda_{2}\sqrt{d})(V_{x}(x) + dV_{y}(x,y) + \bar{\gamma}_{1}\zeta(\widetilde{\mathcal{T}}(\mu,\vartheta))|e|^{2})$. Since in this case we transmit at $\tau = T^{*} \leq \widetilde{\mathcal{T}}(\mu,\vartheta)$, then $\zeta(\tau) \geq \zeta(\widetilde{\mathcal{T}}(\mu,\vartheta))$, as $\zeta(\tau)$ is a decreasing function, and we obtain $R(G(q)) \leq (1 + \lambda_{2}\sqrt{d})R(q)$. When $\tau > T^{*}$, it holds that $\bar{\gamma}_{1}|e|^{2} = \sigma \alpha_{1}V_{x}(x)$ in view of (\ref{flow-jump-sets-clock}). Hence, by following similar lines as above, we deduce that
\begin{equation}\label{R-all-D}
\begin{array}{lllll}
R(G(q)) & \leq & (1+\lambda\sqrt{d})R(q),
\end{array}
\end{equation}
where $\lambda = \max\{\lambda_{2}, (\lambda_{1}+\lambda_{2})\sigma\alpha_{1}\}$. Thus, (\ref{R-all-D}) holds for all $q\in D$ (since $\lambda_{2} \leq \lambda$).

Finally, we use similar arguments as in Proposition 3.29 in \cite{Teel} to conclude. In view of (\ref{eq: thm2-R-flow}) and (\ref{R-all-D}), the property (3.10) in Proposition 3.29 holds with $\lambda_{c} = -\mu$ and $e^{\lambda_{d}} = (1 + \lambda\sqrt{d})$. Let $\psi > 0$ and $(t,j)\in\dom \phi$. To satisfy the last condition of Proposition 3.29, we need to show that $\ln (1 + \lambda\sqrt{d})j-\mu t\leq -\psi(t+j)$. Since $j \leq \frac{t}{T^{*}}$ in view of (\ref{flow-jump-sets-clock}), it suffices to show that $\ln (1 + \lambda\sqrt{d})\frac{t}{T^{*}}-\mu t\leq -\psi(t+\frac{t}{T^{*}})$ which is equivalent to $(\ln (1 + \lambda\sqrt{d})+\psi)\frac{t}{T^{*}} \leq (\mu-\psi)t$, i.e. $\psi(\frac{1}{T^{*}}+1) \leq \mu - \ln (1 + \lambda\sqrt{d})\frac{1}{T^{*}}$. Hence, we take $d \leq (\frac{e^{\mu T^{*}} - 1}{\lambda})^{2}$ which ensures that $\mu - \ln (1 + \lambda\sqrt{d})\frac{1}{T^{*}}>0$ (it then suffices to take $\psi \in (0,\frac{\mu - \ln (1 + \lambda\sqrt{d})\frac{1}{T^{*}}}{\frac{1}{T^{*}}+1})>0$). As a result, like in the proof of Proposition 3.29 in \cite{Teel}, we obtain, for all $(t,j) \in \dom \phi$
\begin{equation}\label{eq: thm2-R-phi}
  R(\phi(t,j)) \leq e^{-\psi(t+j)}R(\phi(0,0)).
\end{equation}
By using Assumptions \ref{ass: assumption-Lyapunov-Vx}-\ref{ass: assumption-Lyapunov-Vy} and the fact that $\zeta(\tau) \in [\vartheta,\vartheta^{-1}]$, we deduce from (\ref{eq: thm2-R-phi}) that (\ref{eq-thm-clock}) holds.
Let $\phi=(\phi_{x},\phi_{y},\phi_{e},\phi_{\tau})$ be a maximal solution to (\ref{hybrid-model})-(\ref{flow-jump-sets-clock}). We note  that $\phi$ is non-trivial by using similar arguments as in the proof of Theorem \ref{thm: practical-stability}.
In view of (\ref{eq: thm2-R-phi}), $\phi_{x}$ and $\phi_{y}$ cannot explode in finite time. Since $\phi_{e}(t,j) = \phi_{x}(t_{j},j) - \phi_{x}(t,j)$ for any $(t_{j},j), (t,j) \in \dom \phi$ and $j \geq 1$, $\phi_{e}$ cannot explode in finite time. The same conclusion holds for $\phi_{\tau}$ in view of its dynamics, see (\ref{hybrid-model}). Hence, $\phi$ cannot explode in finite-time. In addition, $G(D) \subset C$. As a consequence, $\phi$ is complete according to Proposition 6.10 in \cite{Teel}. \hfill $\Box$

\textbf{Proof of Proposition \ref{prop: glob-lip}.}
Let $\bar{L}$ be a common Lipschitz constant for the functions $f, g, k$ and $h$, which exists in view of items (1), (2) of Proposition \ref{prop: glob-lip}.
\begin{itemize}
  \item Assumption \ref{ass: assumption-Lyapunov-Vx}: In view of (\ref{non-linear-slow}) and item (3) of Proposition \ref{prop: glob-lip}, we have, for all $(x,e) \in \mathbb{R}^{2n_{x}}$, $\frac{\partial V_{x}}{\partial x}f_{s}(x,e)\!\leq\! -\bar{\alpha}_{1}V_{x}(x)\!+\!2|x||P_{1}||f_{s}(x,e)\!-\!f_{s}(x,0)|$. Since $f$ is globally Lipschitz, $|f_{s}(x,e) - f_{s}(x,0)| \leq \bar{L}|e|$. As a consequence, $\frac{\partial V_{x}}{\partial x}f_{s}(x,e) \leq -\bar{\alpha}_{1}V_{x}(x) + 2\bar{L}|P_{1}||x||e|$. Using the fact that $2\bar{L}|P_{1}||x||e| \leq \frac{1}{2}\bar{\alpha}_{1}\lambda_{\min}(P_{1})|x|^{2} + 2(\bar{\alpha}_{1}\lambda_{\min}(P_{1}))^{-1}\bar{L}^{2}|P_{1}|^{2}|e|^{2}$ and the fact that $\lambda_{\min}(P_{1})|x|^{2} \leq V_{x}(x) \leq \lambda_{\max}(P_{1})|x|^{2}$, since $P_{1}$ is positive definite and symmetric, it holds that
  \begin{align*}
    \frac{\partial V_{x}}{\partial x}\!f_{s}(x,e) &\!\leq\!-\tfrac{\bar{\alpha}_{1}}{2}V_{x}(x)\!+\! 2\bar{L}^{2}|P_{1}|^{2}(\bar{\alpha}_{1}\lambda_{\min}(\!P_{1}))^{-1}|e|^{2}.
  \end{align*}
  Hence Assumption \ref{ass: assumption-Lyapunov-Vx} holds with $\alpha_{1} = \frac{\bar{\alpha}_{1}}{2}$ and $\gamma_{1}(s)=2\bar{L}^{2}|P_{1}|^{2}(\bar{\alpha}_{1}\lambda_{\min}(P_{1}))^{-1} s^{2}$ for $s \geq 0$.

  \item Assumption \ref{ass: assumption-Lyapunov-Vy} follows directly from item (3) of Proposition \ref{prop: glob-lip}.

  \item Assumption \ref{ass: assumption-Lyapunov-Interconnection-terms}: In view of items (1), (3) of Proposition \ref{prop: glob-lip} and since $\lambda_{\min}(P_{1})|x|^{2} \leq V_{x}(x)$ and $\lambda_{\min}(P_{2})|y|^{2} \leq V_{y}(x,y)$, it holds that, for all $(x,y,e) \in \mathbb{R}^{2n_{x} + n_{y}}$
  \begin{align}
    \frac{\partial V_{x}}{\partial x}\left[ f_{x}(x,\!y,\!e)\!-\!f_{s}(x,\!e)\right]\leq 2\bar{L}|P_{1}||x||y|.
  \end{align}
   Thus, the first condition of Assumption \ref{ass: assumption-Lyapunov-Interconnection-terms} is verified with $\beta_{1} = 2\bar{L}|P_{1}|(\lambda_{\min}(P_{1})\lambda_{\min}(P_{2}))^{-1/2}$. On the other hand, in view of items (1), (3) of Proposition \ref{prop: glob-lip} and using the fact that $|e||y| \leq |y|^{2} + |e|^{2}$ and using that $f_{x}(0,0,0)=0$ since the origin of system (\ref{non-linear-slow}) is asymptotically stable in view of (\ref{eq: prop1-Vx-dot}), it holds that, for all $(x,y,e) \in \mathbb{R}^{2n_{x} + n_{y}}$
  \begin{align*}
    &\left[\!\frac{\partial V_{y}}{\partial x}\!-\!\frac{\partial V_{y}}{\partial y}\frac{\partial h}{\partial x}\!\right]f_{x}(x,y,e) \leq 2|P_{2}||y||\frac{\partial h}{\partial x}||f_{x}(x,y,e)| \nonumber\\
                                            &\hspace{0.8cm} \leq \beta_{2}\sqrt{V_{x}(x)V_{y}(x,y)}\!+\!\beta_{3}V_{y}(x,y) + 2\bar{L}^{2}|P_{2}||e|^{2},
  \end{align*}
  where $\beta_{2} = 2\bar{L}^{2}|P_{2}|(\lambda_{\min}(P_{1})\lambda_{\min}(P_{2}))^{-1/2}$ and $\beta_{3} = 4\bar{L}^{2}|P_{2}|(\lambda_{\min}(P_{2}))^{-1}$. Hence, the second condition of Assumption \ref{ass: assumption-Lyapunov-Interconnection-terms} holds with $\gamma_{2}(s)=2\bar{L}^{2}|P_{2}|s^{2}$ for $s\geq 0$. The third condition is satisfied with $L=\bar{\alpha}_{1}\lambda_{\min}(P_{1})|P_{2}||P_{1}|^{-2}$.

  \item Assumption \ref{ass: Vy-leq-lambdaV}: We denote $h_{xe}(x,e):= h(x,k(x+e))$ and $h_{x}(x):= h(x,k(x))$. In view of (\ref{y-jumps}) and the definition of $V_{y}$,
  \begin{align}
    V_{y}&(x, h_{y}(x,y,e)) = h_{y}^{T}(x,y,e)P_{2}h_{y}(x,y,e) \nonumber\\
                           &\leq V_{y}(x,y) + |P_{2}||h_{xe}(x,e) - h_{x}(x)|^{2} \nonumber\\
                           &\quad + 2|P_{2}||y||h_{xe}(x,e) - h_{x}(x)|.
  \end{align}
  Since $h$ is globally Lipschitz, it holds that $|h_{xe}(x,e) - h_{x}(x)| \leq \bar{L}|e|$. As a consequence,
  \begin{align*}
    V_{y}(x, h_{y}(x,y,e))\!\leq\!V_{y}(x,y)\!+\!\bar{L}^{2}|P_{2}||e|^{2}\!+\!2\bar{L}|P_{2}||y||e|
  \end{align*}
  and Assumption \ref{ass: Vy-leq-lambdaV} is satisfied with $\lambda_{1} = \frac{1}{2}\bar{\alpha}_{1} \lambda_{\min}(P_{1}) \times$ $|P_{2}| |P_{1}|^{-2}$ and $\lambda_{2} = (\bar{\alpha}_{1}\lambda_{\min}(P_{1})|P_{2}|^{2})^{1/2} \times$\\ $(\lambda_{\min}(P_{2})|P_{1}|^{2})^{-1/2}$.

  \item Assumption \ref{ass: clock}: In view of (\ref{hybrid-model})-(\ref{flow_jump_map}) and item (1) of Proposition \ref{prop: glob-lip} and using that $f_{x}(0,0,0) = 0$, it holds that, for all $(x,y)\in \mathbb{R}^{n_{x}+n_{y}}$ and for almost all $e \in \mathbb{R}^{n_{x}}$
  \begin{align}
    \langle \nabla|e|, -f_{x}(x,y,e)\rangle \leq \bar{L}(|x| + |y| + |e|).
  \end{align}
  Hence, Assumption \ref{ass: clock} is verified with $M = \bar{L}$ and $N = \bar{L}\max\{(\lambda_{\min}(P_{1}))^{-1/2}, (\lambda_{\min}(P_{2}))^{-1/2}\}$.  \hfill $\Box$
  \end{itemize}

\end{document}